\documentclass[a4paper]{jpconf}
\usepackage{graphicx}
\usepackage{hyperref}

\def\asec{\ifmmode ^{\prime\prime}\else$^{\prime\prime}$\fi}

\def\it{\sl}
\def\degs{\ifmmode ^{\circ}\else$^{\circ}$\fi}
\def\amin{\ifmmode ^{\prime}\else$^{\prime}$\fi}
\def\asec{\ifmmode ^{\prime\prime}\else$^{\prime\prime}$\fi}
\def\fss{\hbox{$.\!\!^{\rm s}$}}        
\def\farcs{\hbox{$.\!\!^{\prime\prime}$}}  
\def\h{$^{\rm h}$}
\def\m{$^{\rm m}$}

\def\degs{\ifmmode ^{\circ}\else$^{\circ}$\fi}
\def\amin{\ifmmode ^{\prime}\else$^{\prime}$\fi}
\def\farcm{\hbox{$.\mkern-4mu^\prime$}}
\def\eqalign#1{\null\,\vcenter{\openup1\jot \m@th
   \ialign{\strut\hfil$\displaystyle{##}$&$\displaystyle{{}##}$\hfil
   \crcr#1\crcr}}\,}
\def\psr{J1932}
\def\chan{\textit{Chandra}}
\def\suz{\textit{Suzaku}}
\def\swift{\textit{Swift}}

\usepackage{xcolor}

\usepackage[normalem]{ulem}  

\begin{document}
\title{Studying the $\gamma$-ray pulsar J1932+1916 and its pulsar wind nebula with \chan}
\author{O D Medvedev$^1$, 
A V Karpova$^1$,
Yu A Shibanov$^1$,
D A Zyuzin$^1$ and \\
G G Pavlov$^2$
}

\address{$^1$Ioffe Institute, Politekhnicheskaya 26, St. Petersburg, 194021, Russia \\
$^2$Pennsylvania State University, 525 Davey Lab., University Park, PA 16802, USA \\
}

\ead{medvedev.oleg12@gmail.com}

\begin{abstract}
We report on the results of \chan\ X-ray observations of the 
$\gamma$-ray radio-quiet pulsar J1932+1916. 
We confirm the previous detection of the pulsar counterpart and
its pulsar wind nebula in X-rays from low spatial resolution
data obtained by \suz\ and \swift.
The \chan\ data with much better spatial resolution resolved the fine structure of  the nebula in the pulsar vicinity, 
which  can be interpreted as jets bent by the ram pressure. The size of this compact part is about 30\asec, 
and it is surrounded by a weak asymmetric  diffusive emission  extended up to $\approx$2\farcm8.
The X-ray spectra of the pulsar, the compact and extended parts of the nebula  can be described 
by the power-law models with photon indexes of $0.44^{+0.57}_{-0.61}$, $2.34^{+0.95}_{-0.79}$ 
and $2.14^{+0.32}_{-0.31}$, respectively. 
The actual pulsar's X-ray flux is several times weaker than it was obtained with  
\suz\ and \swift\ where it was dominated by the unresolved nebula. 
We discuss possible associations of J1932+1916 with the nearby supernova remnant G54.4$-$0.3.

\end{abstract}


\section{Introduction}

To date \textit{Fermi} $\gamma$-observatory has detected 234 
pulsars 
(see, e.g., \url{https://confluence.slac.stanford.edu/display/GLAMCOG/Public+List+of+LAT-Detected+Gamma-Ray+Pulsars}).
Several dozens of them are radio-quiet, which makes X-ray observations 
particularly important for further studies of their properties. 
  
The $\gamma$-ray radio-quiet pulsar J1932$+$1916 (hereafter \psr) 
was discovered in a `blind search' of the \textit{Fermi} data \cite{Pletsch2013}.
The pulsar has the following parameters:
R.A.=19\h32\m19\fss70(4), Dec.=+19\degs16\amin39\asec(1) 
(hereafter, numbers 
in parentheses are 1$\sigma$ uncertainties in the last digits quoted unless stated otherwise), 
the period $P=208$~ms, the characteristic age $\tau = 35$~kyr, 
the spin-down luminosity $\dot{E}=4\times10^{35}$~erg~s$^{-1}$, 
the magnetic field $B=4.5\times10^{12}$~G
and the $\gamma$-ray flux above 100 MeV 
$G_{100}=7.8(4)\times10^{-11}$ erg cm$^{-2}$ s$^{-1}$.

First X-ray observations of the \psr\ field with the \swift\ and 
\suz\ telescopes  
revealed a possible pulsar counterpart as well as
a presumed pulsar wind nebula (PWN) \cite{karpova2017}.
However, low spatial resolution of the instruments 
did not allow us to study the objects in detail.
To confirm the \psr\ counterpart candidate 
and to clarify the properties of the pulsar$+$PWN system,
we performed new observations with \chan,
 which provides the best spatial resolution in X-rays.
Here we present 
the results of the observations. 

\begin{figure}
\begin{minipage}[h]{0.495\linewidth}
\center{\includegraphics[width=1.0\linewidth, clip]{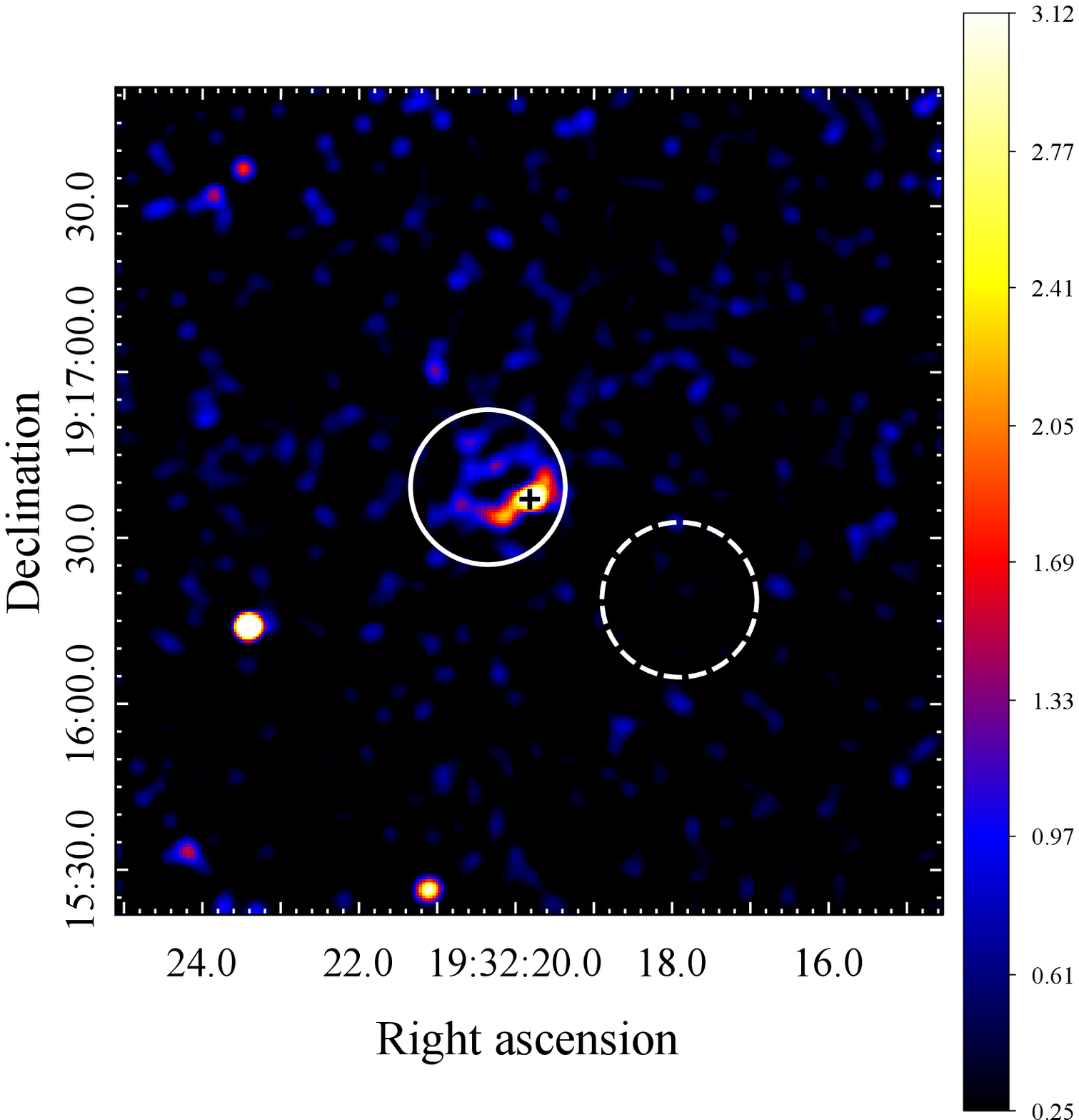}}
\put (-96,134) {\includegraphics[scale=0.22,clip]{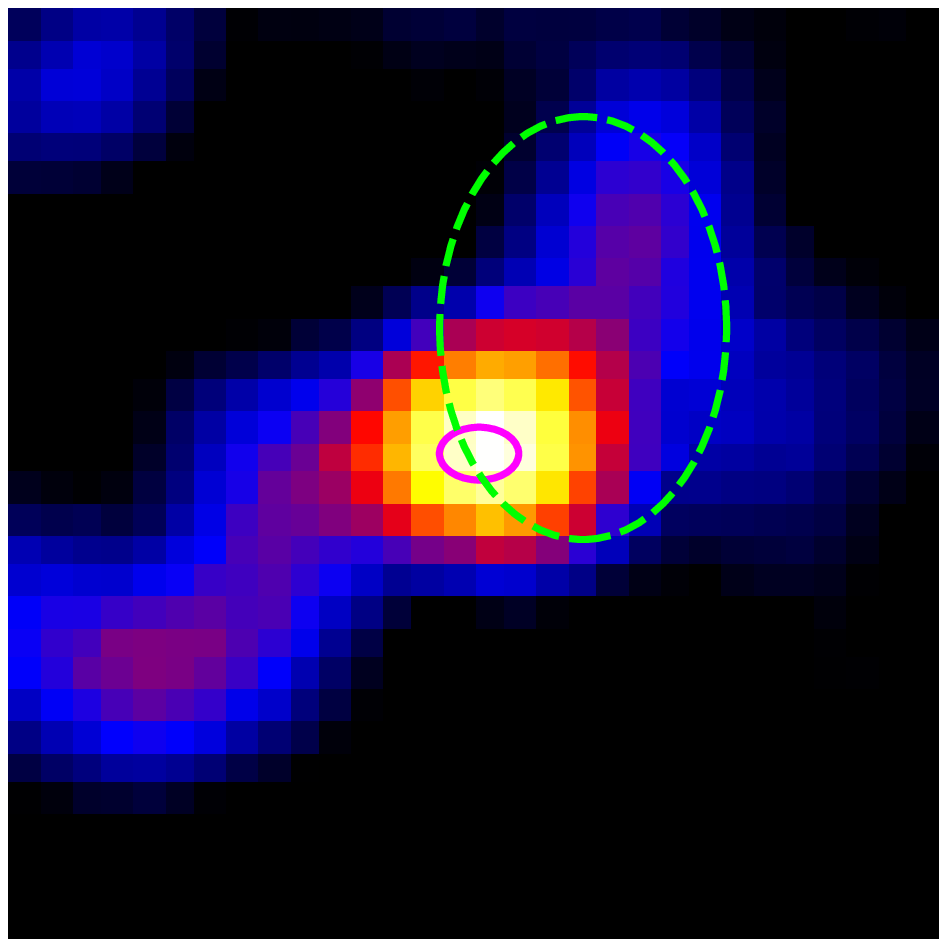}}
\end{minipage}
\begin{minipage}[h]{0.495\linewidth}
\center{\includegraphics[width=1.0\linewidth, clip]{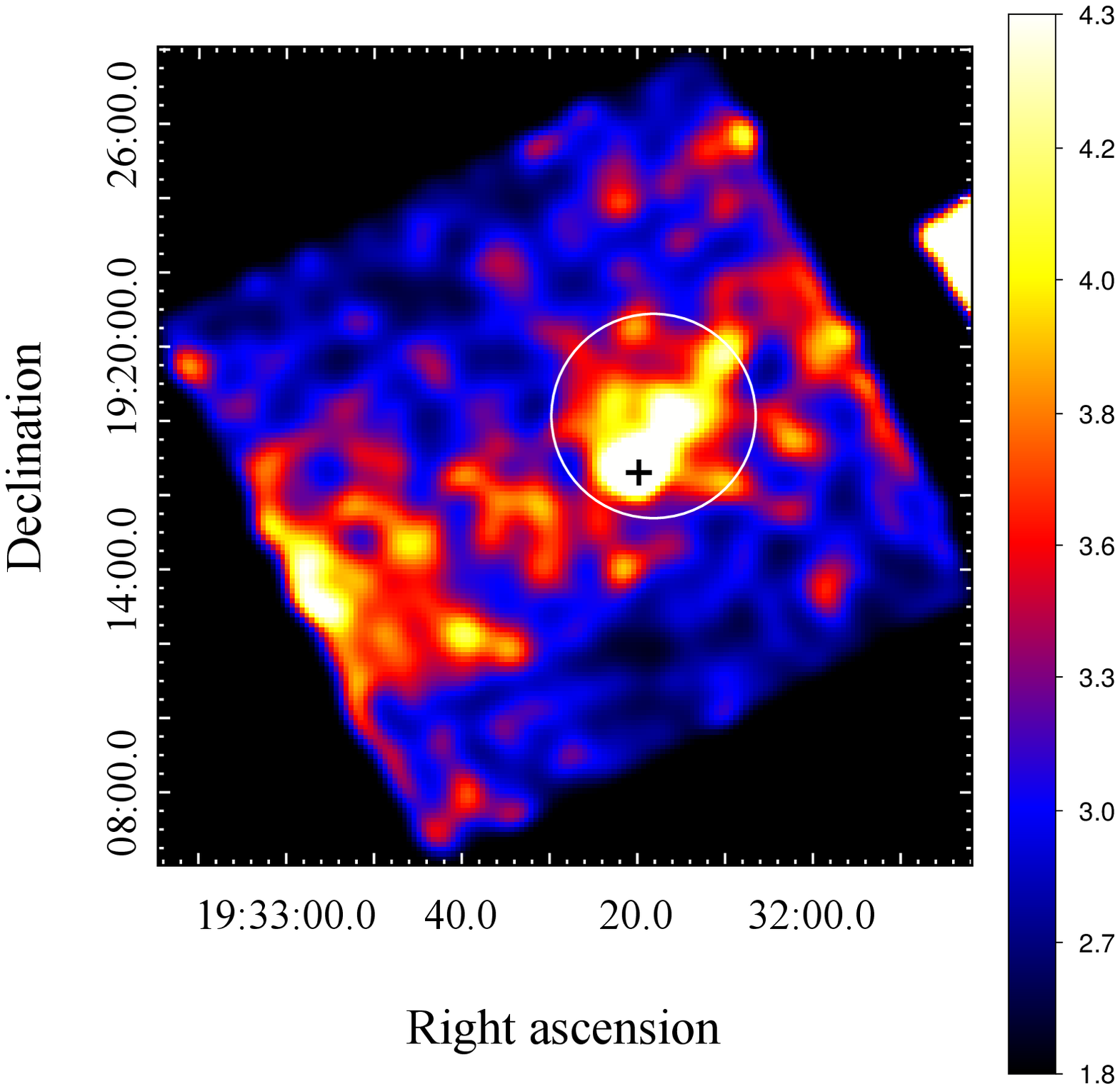}}
\end{minipage}
\caption{\textit{Left}: 2\farcm5$\times$2\farcm5 exposure-corrected \chan\ ACIS-I  
image of the \psr\ field in the 0.5--7 keV band
smoothed with a 6 pixel Gaussian kernel (1 pixel = 0\farcs5). 
The intensity is given in $10^{-8}$ photon~cm$^{-2}$~s$^{-1}$~pixel$^{-1}$.
The likely pulsar counterpart position is shown by the cross.
The solid and dashed circles were used for the extraction 
of the compact PWN and background spectra, respectively.
The internal box show
the zoomed-in 15\asec$\times$15\asec\ region 
around the \psr\ counterpart candidate
smoothed with a 5 pixel Gaussian kernel.
The green dashed and magenta solid ellipses show 3$\sigma$ 
position uncertainties of \psr\ in $\gamma$-rays and 
the presumed X-ray counterpart, respectively. 
The green ellipse combines the \textit{Fermi} 
position uncertainties 
and the \chan\ astrometric accuracy (the latter is estimated 
from 90\% confidence \chan\ aspect solution accuracy of 0\farcs8; 
see \url{http://cxc.cfa.harvard.edu/cal/ASPECT/celmon/}).
\textit{Right}: Exposure-corrected image of the ACIS-I FOV in 
the 0.5--7 keV band smoothed with a 6 pixel Gaussian kernel (1 pixel = 8\asec). 
The intensity is given in $10^{-7}$ photon~cm$^{-2}$~s$^{-1}$~pixel$^{-1}$.
The 2\farcm75-radius circle marks the large-scale nebula emission around \psr.
Extended emission near the south-east edge of the FOV is likely associated with the shell 
of SNR    G54.4$-$0.3. } 
\label{fig:images} 
\end{figure}
\section{X-ray data and imaging}
\label{sec: data}

Two 15-ks \chan\ ACIS-I observations of the \psr\ field were carried out 
on 2018 February 26 and 28 (ObsIDs 20742 and 20986; PI G. Garmire).
The data were reprocessed and analysed using {\tt CIAO v.4.11} package. 
The exposure-corrected image of the pulsar vicinity is shown in the left panel of 
figure~\ref{fig:images} where the data from the two sets were combined.
A point-like source is seen at the $\gamma$-ray position of \psr\
as well as a weak compact nebula 
around it, which resembles two bent tails (or jets).  
The point source position obtained with the \texttt{wavdetect} tool
on the merged image is R.A.=19\h32\m19\fss817(15) and Dec.=+19\degs16\amin37\farcs00(14) 
where errors in brackets are pure statistical uncertainties. 

To search for a large-scale X-ray emission in the pulsar field, 
previously observed with \suz\ \cite{karpova2017},
we created the image where point-like sources were excluded. 
The resulting holes were filled with background
counts taken from annuli regions around the sources using the 
\texttt{dmfilth} tool. The image of the ACIS-I field-of-view (FOV) 
is presented in the right panel of figure~\ref{fig:images}. 
One can see some extended diffuse emission around \psr\ 
which may be a fainter part of its PWN.
Its size is about 5\farcm5. It coincides with the extended source 
(the presumed PWN) detected with \suz. 
The surface brightness of the latter can be described 
by the 2D Gaussian model with FWHM$\sim$4\farcm5 \cite{karpova2017}.
Note that diffuse emission south-east of the pulsar (near the ACIS-I edge)
may be related to the supernova remnant (SNR) G54.4$-$0.3
which was proposed as the \psr\ host remnant \cite{karpova2017}.

\section{Spectral analysis}
\label{sec: spec_an}

The X-ray spectra in both datasets 
are extracted using {\tt specextract} task.
For the \psr, we 
use 
the 2\asec-radius aperture, while the 14\asec-radius circle 
with the pulsar aperture subtracted 
is used 
for the compact PWN (figure~\ref{fig:images}).
The background spectra 
are extracted from the dashed circle. 
For each object, spectra 
are combined to increase signal-to-noise ratio
and 
grouped to ensure at least 1 count per energy bin. 
We obtain 
19 and 83 total source counts from \psr\ and its nebula, respectively.
Due to the low count number, we 
use $C$-statistics \cite{cash1979} to fit the data.
We 
fit 
the compact nebular (CN) 
spectrum containing more source counts in the 0.5--10 keV band 
using \texttt{XSPEC v.12.9.1} and the absorbed power law (PL) model.
We 
obtain
the following best-fit parameters (hereafter all errors correspond to 1$\sigma$ confidence intervals):
the absorbing column density $N_{\rm H}=1.36^{+1.08}_{-0.78}\times10^{22}$ cm$^{-2}$,
the photon index $\Gamma^{\rm CN}=2.34^{+0.95}_{-0.79}$, the unabsorbed flux in the 0.5--8 keV band
$F_X^{\rm CN}=1.0^{+1.8}_{-0.4} \times 10^{-13}$ erg~cm$^{-2}$~s$^{-1}$
and $C=65.7$ (degrees of freedom d.o.f.= 96).
We 
fit 
the \psr\ counterpart candidate spectrum
fixing the column density at the best-fit value for the PWN.
The PL model fit is: 
$\Gamma^{\rm PSR}=0.44^{+0.57}_{-0.61}$, 
unabsorbed flux $F_X^{\rm PSR}=2.2^{+0.6}_{-0.5} \times 10^{-14}$ erg~cm$^{-2}$~s$^{-1}$ 
and $C=14.2$ (d.o.f. = 16).
The black body (BB) model for the pulsar 
suggests 
a temperature of $1.4^{+0.8}_{-0.4}$ keV
which is too large
for the entire surface of a neutron star (NS) or for its hot polar caps
(see e.g. \cite{vigano2013}).

We 
analyze 
the 
extended nebula (EN) spectrum using 
the circle region shown 
in the right panel of figure~\ref{fig:images} 
to extract the spectrum (the 14\asec-radius circle used 
for the compact PWN analysis was excluded as well as point-like sources). 
We grouped the spectrum to ensure 20 counts per energy bin
and used $\chi^2$-statistics.
We 
fit 
this spectrum fixing the column density at the best-fit value for   the compact PWN. 
The PL model results in $\Gamma^{\rm EN} = 2.14^{+0.32}_{-0.31}$, unabsorbed flux $F_{X}^{\rm EN} = 8.3^{+1.3}_{-1.1} \times 10^{-13}$ erg~cm$^{-2}$~s$^{-1}$
and $\chi^2=228$ (d.o.f. = 199).  


\section{Discussion and conclusions}


\begin{figure}
\begin{minipage}[h]{0.33\linewidth}
\center{\includegraphics[width=1.0\linewidth, clip]{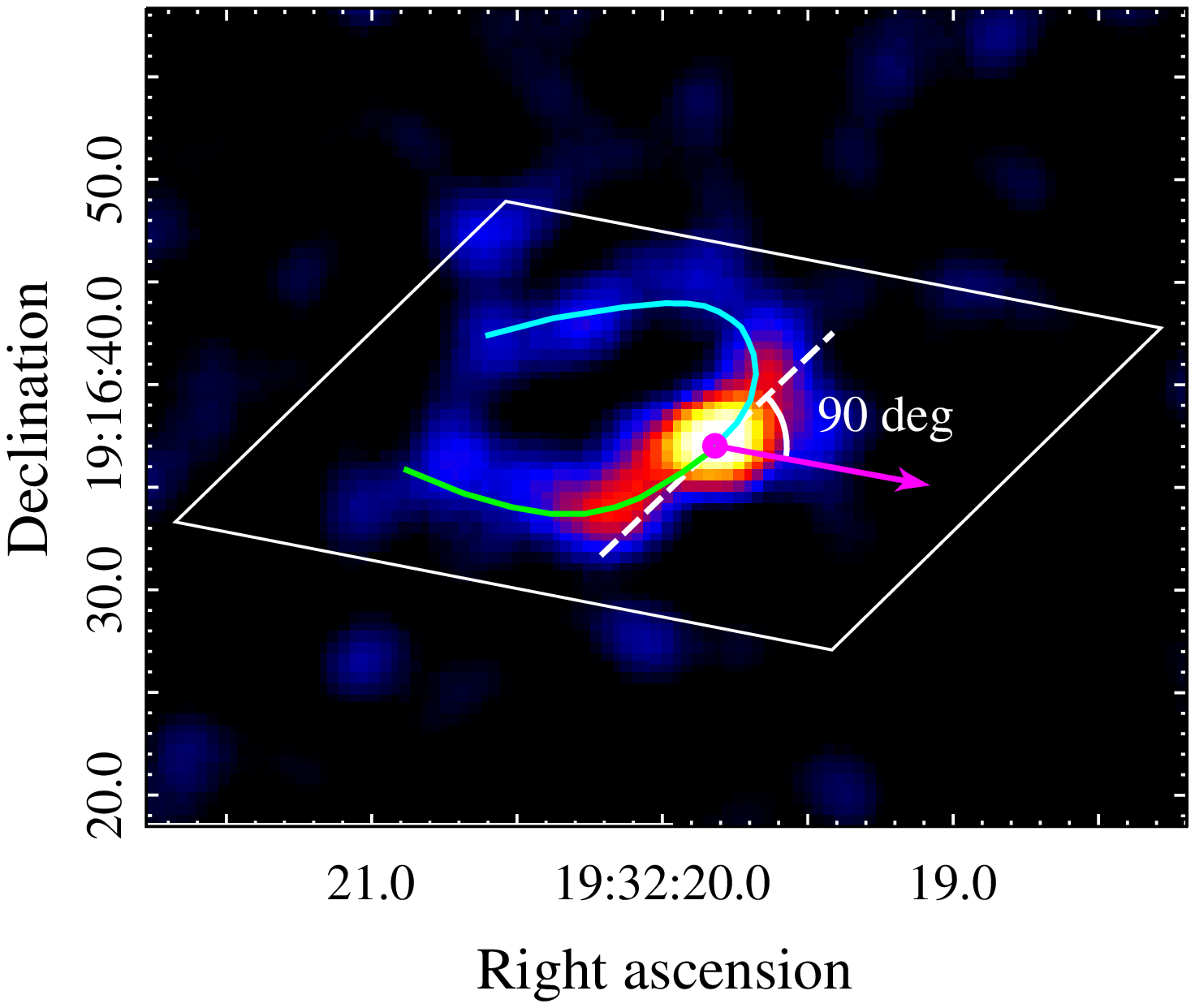}}
\end{minipage}
\begin{minipage}[h]{0.33\linewidth}
\center{\includegraphics[width=1.0\linewidth, clip]{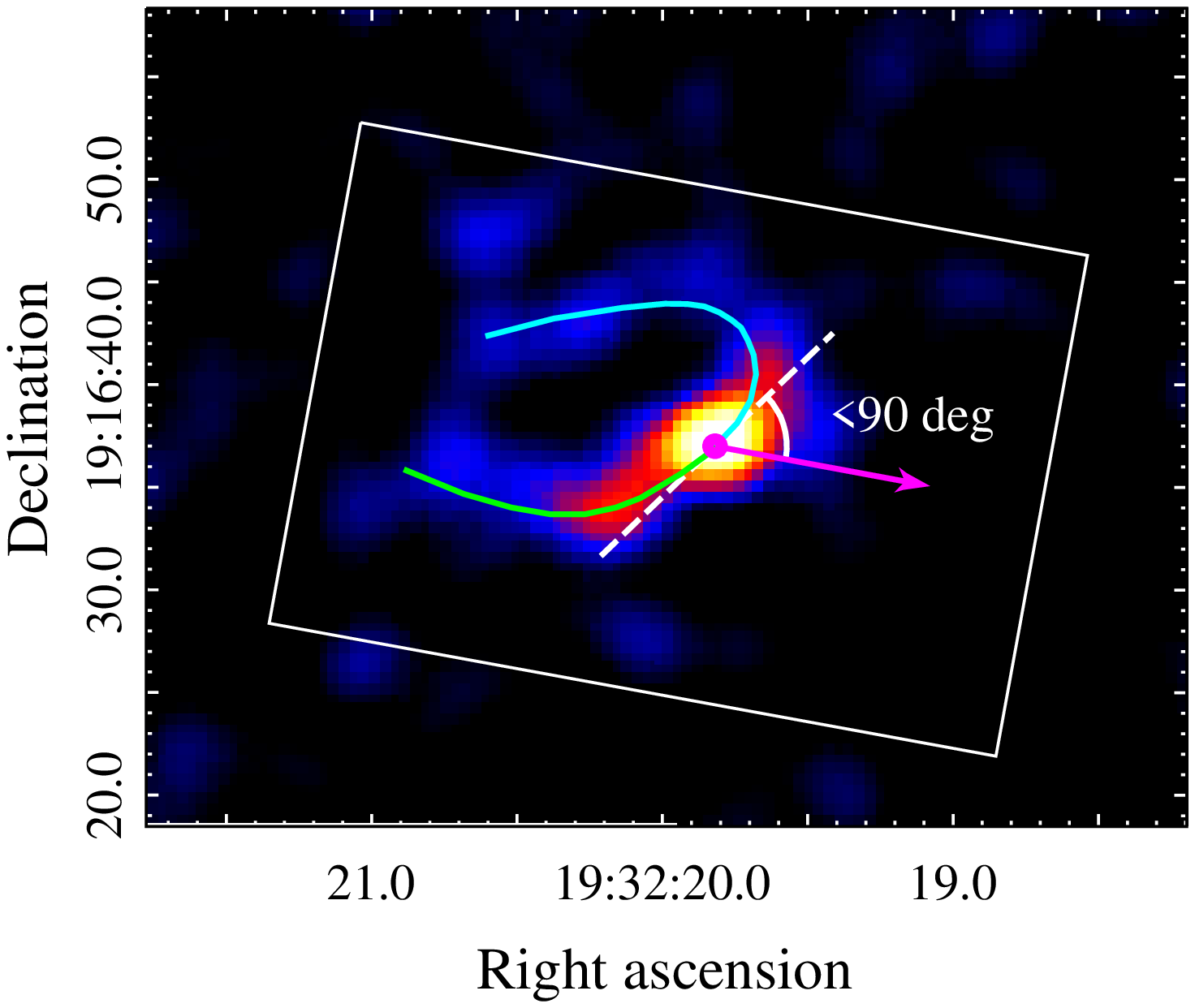}}
\end{minipage}
\begin{minipage}[h]{0.33\linewidth}
\center{\includegraphics[width=1.0\linewidth,clip]{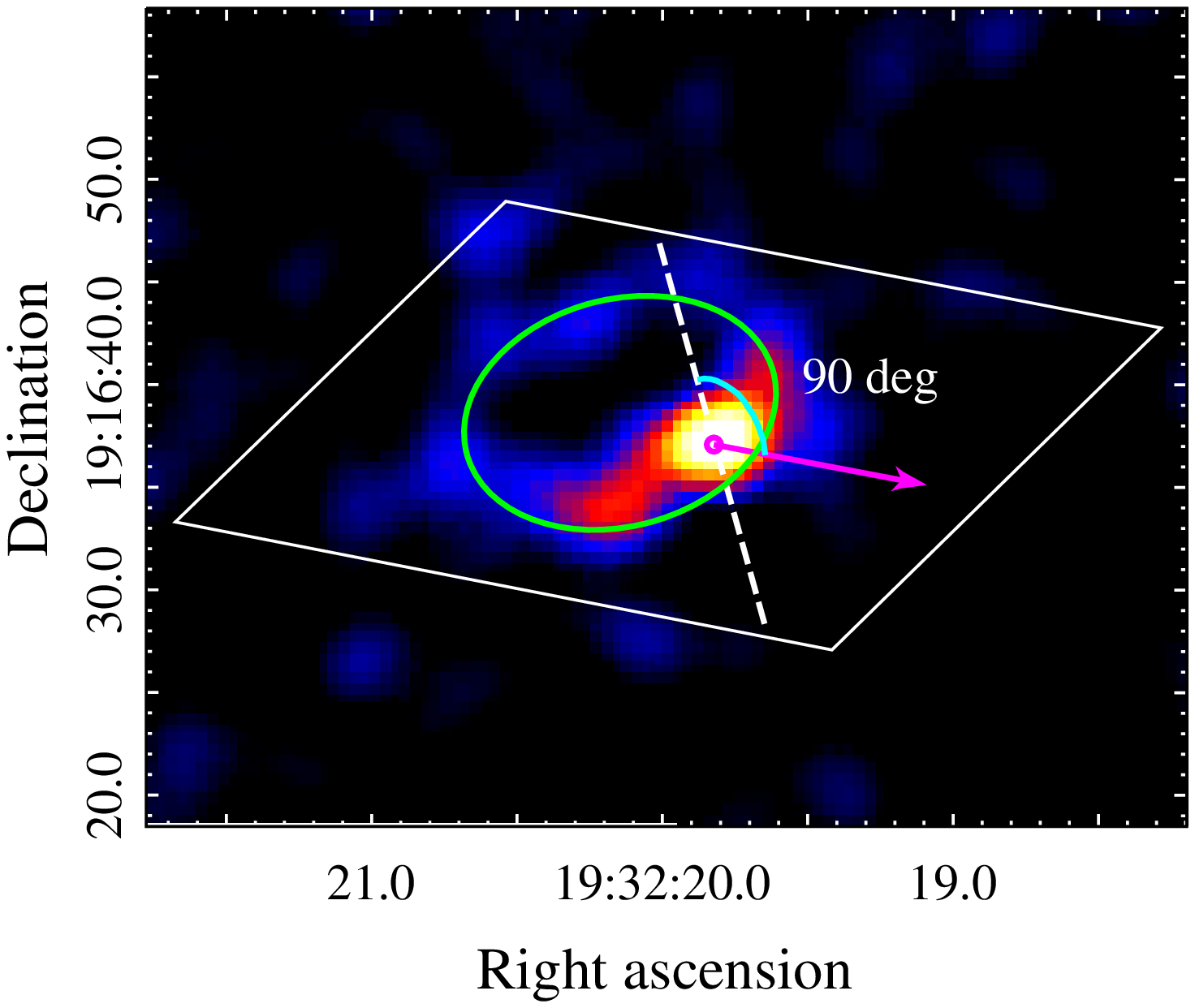}}
\end{minipage}
\caption{50\asec$\times$40\asec ACIS-I images of the pulsar field in the 0.5--7 keV band
encapsulating the compact nebula part. 
Possible geometries of the system 
are shown.
The assumed pulsar's 
proper motion (p.m.)
direction and its spin axis are shown 
by the magenta arrow and the 
white dashed
line, respectively.
White tetragons indicate the plane of the system.
In the left and middle panels the green and cyan lines 
represent jets bent by the ram pressure.
In the right panel the green ellipse shows the equatorial outflow (torus)
shifted by the ram pressure. See text for details.
}
\label{fig:geom}
\end{figure}

\begin{figure}
\begin{minipage}[h]{1.0\linewidth}
\center{\includegraphics[width=0.5\linewidth, clip]{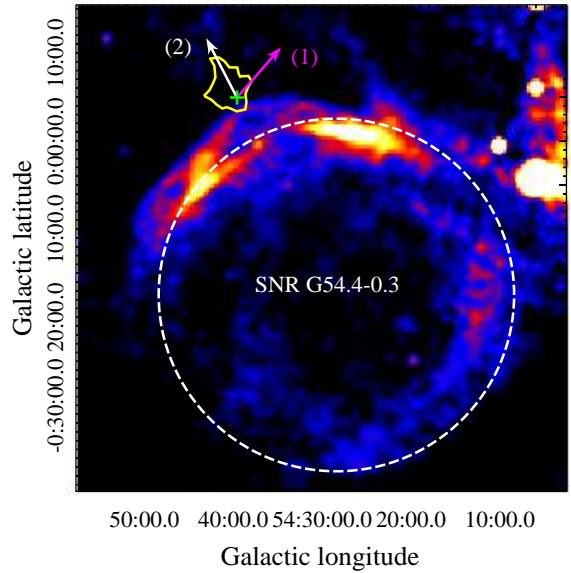}}
\end{minipage}
\caption{Image of the SNR G54.4$-$0.3 obtained from the VGPS.
The \psr\ X-ray counterpart position is marked by the green cross.
The magenta arrow (1) indicates the pulsar p.m. 
as shown in figure~\ref{fig:geom}. 
The white arrow (2) shows the pulsar's p.m. in the case of its
association with G54.4$-$0.3.
The yellow contour corresponds to the boundary of the large-scale PWN emission around \psr\ seen in the right panel 
of figure~\ref{fig:images}.}
\label{fig:snr}
\end{figure}
  
The \psr\ X-ray counterpart flux measured from 
\chan\ data 
is by a factor of 5  smaller than that  obtained in \cite{karpova2017}. 
It is clear now that the latter was dominated  
by the compact PWN structures 
around the pulsar which were not resolved by \suz\ and \swift.
The ratio of the pulsar fluxes in $\gamma$-rays and X-rays is log$(G_{100}/F_X^{\rm PSR})\approx 3.5$ which is in agreement with the value 
typical for the radio-quiet $\gamma$-ray pulsar population \cite{abdo2013ApJS}.

The derived  photon indices indicate that the PWN emission appears to be softer 
than that of the pulsar.  
The column density $N_{\rm H}$ is in agreement with the value
obtained in \cite{karpova2017} though it is
too uncertain to better constrain the distance to the pulsar. 
The unabsorbed flux from the extended nebula is compatible with
the value obtained for the entire PWN seen in the \suz\ data \cite{karpova2017}.


The \psr\ compact nebula may be similar to the `lateral tails' 
observed in the 
PWN
of PSRs J0633+1746 (Geminga) and J1509$-$5850
\cite{kargaltsev2017,posselt2017,klinger2016}.
These 
tails could 
represent a limb-brightened shell formed by the ram pressure
of the ISM interacting with the pulsar wind.
The shape of such a shell may be a body of rotation with the symmetry axis along the
pulsar's 
p.m. direction. A deficit of diffuse emission between the tails in this case
can be caused by magnetic field dependence of the azimuthal angle
around the shell axis \cite{posselt2017}. 
This explanation seems to be implausible for \psr\ 
due to observed asymmetry of the compact nebula.
However, the axially asymmetric shell is expected 
if the pulsar's p.m. direction and spin axis are not aligned, 
and the pulsar wind is concentrated in the equatorial plane \cite{posselt2017}. 
For example, if the spin axis is perpendicular to the p.m. direction
and the plane of the sky, two bent streams without emission between them
could be observed.

There is another interpretation of `lateral tails' as jets or 
an equatorial torus bent back by the ram pressure \cite{kargaltsev2017,posselt2017,klinger2016}.
We show possible geometries in figure~\ref{fig:geom}.
The system asymmetry can be explained as follows:
the plane of the system defined by the p.m. and spin vectors can be not perpendicular 
to the line of sight (figure~\ref{fig:geom}, left and right) or
the spin axis can be not perpendicular to the pulsar's 
p.m. direction (figure~\ref{fig:geom}, middle).

All the geometries for \psr\ system shown in figure~\ref{fig:geom} 
assume the same direction of the pulsar p.m. 
As was noted above, \psr\ can be associated with 
the SNR G54.4$-$0.3 \cite{karpova2017}. 
The radio image of the latter obtained from the VLA Galactic Plane Survey (VGPS) 
\cite{stil2006} is presented in figure~\ref{fig:snr}.
The association of \psr\ and G54.4$-$0.3 requires the pulsar's p.m.
direction (the white  arrow in figure~\ref{fig:snr}) very different from that one indicated in figure~\ref{fig:geom}.
Thus, the geometries assumed in figure~\ref{fig:geom} reject the association.

Recently, a $\sim$5\asec-radius ring structure was detected around
the presumed pulsar CXOU J061705.3+222127 (J0617) in the SNR IC 443 \cite{swartz2015}.
The center of the ring is offset by 2\farcs7 from the J0617 position.
This may be explained either as an intrinsic azimuthal asymmetry provided by 
nonuniform circumstellar medium or as a geometric effect 
due to a 
 shift of the ring  from the equatorial plane \cite{swartz2015}. 
The same situation can take place for \psr.
In this case its p.m. direction is unclear, and
we cannot exclude 
G54.4$-$0.3 as the host remnant, presuming    
 the  p.m.  direction of \psr\ along the white arrow in figure~\ref{fig:snr}.     
The large scale  PWN (right panel of figure~\ref{fig:images}),  
whose extent is shown by the yellow contour in figure~\ref{fig:snr}, 
appears to be  stretched 
along this arrow, suggesting some protrusion ahead of the pulsar.

Deeper observations are necessary to clarify the nature 
of the small- and large-scale PWN structures around \psr.
Detection of the proper motion and X-ray pulsations from \psr\ could help to 
understand geometry of the system and its association with G54.4$-$0.3.

\ack{GGP acknowledges support from 
the ACIS Instrument Team contract SV4-74018 issued by the Chandra X-ray Observatory Center, 
which is operated by the Smithsonian Astrophysical Observatory for 
and on behalf of NASA under contract NAS8-03060.
The ACIS Guaranteed Time Observations included here were selected by the
ACIS Instrument Principal Investigator, Gordon P.\ Garmire, of the
Huntingdon Institute for X-ray Astronomy, LLC, which is under contract
to the Smithsonian Astrophysical Observatory; Contract SV2-82024.
For figure~\ref{fig:snr} we used the data from the VGPS survey 
conducted by the National Radio Astronomy Observatory (NRAO) instruments. 
NRAO is a facility of the National Science Foundation 
operated under cooperative agreement by Associated Universities, Inc.
DAZ thanks Pirinem School of Theoretical Physics for hospitality.}


\section*{References}

\bibliography{ref1932}

\end{document}